\documentclass[reprint, superscriptaddress, amsmath, amssymb, aps, showkeys]{revtex4-1}

\usepackage{graphicx}
\usepackage{booktabs}
\usepackage{url}
\usepackage{natbib}

\begin{document}

\title{Constructing multi-level urban clusters based on population distributions and interactions}

\author{Wenpu Cao}
\affiliation{Institute of Remote Sensing and Geographic Information Systems, School of Earth and Space Sciences, Peking University, Beijing 100871, China}

\author{Lei Dong}
\email{Corresponding author: leidong@pku.edu.cn}
\affiliation{Institute of Remote Sensing and Geographic Information Systems, School of Earth and Space Sciences, Peking University, Beijing 100871, China}

\author{Ying Cheng}
\affiliation{State Key Laboratory of Media Convergence Production Technology and Systems, Beijing, China}

\author{Lun Wu}
\affiliation{Institute of Remote Sensing and Geographic Information Systems, School of Earth and Space Sciences, Peking University, Beijing 100871, China}

\author{Qinghua Guo}
\affiliation{Institute of Remote Sensing and Geographic Information Systems, School of Earth and Space Sciences, Peking University, Beijing 100871, China}

\author{Yu Liu}
\affiliation{Institute of Remote Sensing and Geographic Information Systems, School of Earth and Space Sciences, Peking University, Beijing 100871, China}
 
\begin{abstract}
A city (or an urban cluster) is not an isolated spatial unit, but a combination of areas with closely linked socio-economic activities. However, so far, we lack a consistent and quantitative approach to define multi-level urban clusters through these socio-economic connections. Here, using granular population distribution and flow data from China, we propose a bottom-up aggregation approach to quantify urban clusters at multiple spatial scales. We reveal six `phases' (i.e., levels) in the population density-flow diagram, each of which corresponds to a spatial configuration of urban clusters from large to small. Besides, our results show that Zipf's law appears only after the fifth level, confirming the spatially dependent nature of urban laws. Our approach does not need pre-defined administrative boundaries and can be applied effectively on a global scale.
\end{abstract}


\maketitle

\section{Introduction}
A city (or an urban cluster) is an integrated area with a dense population and intensive internal socio-economic connections. It includes not only the core urban area but also the surrounding areas that interact closely with the core \citep{berry_metropolitan_1969, dash_nelson_economic_2016}. However, in many countries, administrative divisions are still the most common way of delineating urban areas, which poses many urban planning and transportation problems \citep{ma_functional_2020}. For example, in China, many people work in Beijing but live in Sanhe (a small neighboring county close to Beijing) due to the low cost of living and relatively convenient transportation \citep{bai_view_2016}. Those people who commute between Sanhe and Beijing contribute to Beijing's economic growth and have demands for Beijing's public services, but they are not part of Beijing's resident population according to the current definition of the city's boundaries.

Realizing that the administrative boundaries of cities are problematic, many studies turn to delineating urban areas through data-driven ways. These efforts include the identification of contiguous urban areas with a high density of human activity, i.e., density-based methods \citep{imhoff_technique_1997, rozenfeld_laws_2008, rozenfeld_area_2011, zhou_cluster-based_2014, arcaute_cities_2016}. For example, the threshold methods define areas that exceed a certain density of population, infrastructure (e.g., road networks), or socio-economic activity (e.g., nighttime light) as urban areas \citep{rozenfeld_laws_2008, jiang_zipfs_2011, cao_quantifying_2020, baragwanath_detecting_2021}. Some machine learning methods extract impervious surfaces or high-density built-up areas as urban areas from remote sensing data \citep{schneider_mapping_2010, galdo_identifying_2021}. 

With the rapid development of transportation and communication technology, the flows of people, goods, and information between areas are becoming increasingly frequent. From a flow perspective, a city can be defined in terms of its core urban areas and surrounding areas with intensive socio-economic interactions, known as functional urban areas (FUAs) \citep{hall_looking_2009, noauthor_eu-oecd_2019, chen_delineating_2022}. Some developed countries have official definitions for FUAs, such as the metropolitan statistical areas (MSAs) in the United States, which encompass areas with high-density populations and the surrounding commuting zones \citep{berry_metropolitan_1969}. Some recent studies have used similar flow-based methods to quantify urban areas in the UK \citep{arcaute_constructing_2015} and France \citep{cottineau_diverse_2017, cottineau_defining_2019}. However, most previous works to identify FUAs need detailed census data or commuting survey data, which are difficult to obtain in developing countries. Meanwhile, the choice of population and commuting thresholds is arbitrary in most methods, making it difficult to compare urban clusters across different countries and periods \citep{wu_zipfs_2018}.

Another piece of the puzzle that has been overlooked by previous flow-based methods is the multi-level structure of urban systems. For example, the flow-based method by the OECD defines urban clusters on only one spatial level \citep{noauthor_eu-oecd_2019}, whereas urban systems have multiple levels, according to previous studies \citep{batty_new_2013, arcaute_cities_2016, alessandretti_scales_2020}. Although several data-driven approaches are proposed to derive urban hierarchies based on percolation theory \citep{arcaute_cities_2016} or community detection \citep{hong_hierarchical_2019, jia_delineating_2021}, robust methods to delineate multi-scale urban clusters and reveal the different levels of urban systems require further research.

With the development of information and communication technology, mobile phone data have shown great potential in mapping dynamic population distributions \citep{deville_dynamic_2014, chen_fine-grained_2018} and flow interactions \citep{kung_exploring_2014, hadachi_unveiling_2020}. These advances in mobile phone data allow us to quantify multi-scale urban clusters in terms of both population distributions and flow interactions \citep{deville_dynamic_2014, hadachi_unveiling_2020, chen_delineating_2022}. In addition, city science has shown that urban systems are typically complex systems and obey some universal patterns, such as Zipf's law \citep{zipf_human_1949, gabaix_zipfs_1999} and allometric growth \citep{nordbeck_urban_1971, bettencourt_growth_2007}, which guide us to study effective multi-level urban clusters. 

In this paper, we propose a bottom-up approach to quantify multi-level urban clusters based on the population distributions and flows inferred from mobile phone data. We first traverse all population density thresholds and aggregate spatially contiguous populated units into separate clusters under each threshold. This process is similar to previous density-based methods \citep{rozenfeld_laws_2008, rozenfeld_area_2011}, and we refer to these individual clusters as population clusters. We then link population clusters to form urban clusters based on different flow ratio thresholds. Through percolation analysis, we obtain six `phases' in the population density-flow phase diagram, each of which corresponds to one level of urban systems. Results show that Zipf's law only appears after the fifth level. Our research proposes an effective method to characterize the population distributions and interactions of cities and to quantitatively understand urban clusters on different levels.

\section{Data}
The population distribution and flow data that we used in this work are derived from an anonymous mobile phone dataset provided by a large location service provider from China in 2015 \citep{cao_quantifying_2020, dong_measuring_2017, dong_universality_2022}. Individual's home and work locations are detected by a machine learning method from the user's stay points; see \citet{dong_universality_2022} for details. The population distribution is calculated based on the number of users whose homes are located within each $0.5^\prime \times 0.5^\prime$ (approximately $1km$ at the equator) grid. The commuting flow data aggregate the number of home-work connections between any two $0.5^\prime \times 0.5^\prime$ grids.

Different location service providers have different market shares in different regions. Therefore, we need to assess whether the population data derived from one single service provider can represent the real population dynamics in China. To assess the accuracy of the population distribution inferred from the mobile phone data, we compare the estimated data with the micro-census data of 2015 at the prefecture level. The correlation coefficient is $0.9$, indicating that the estimated population distribution is consistent with the census. Note that the mobile phone dataset covers approximately 100 million people out of a total population of 1.4 billion in China. Thus, all reported population-related values are scaled by $13.9$ (total population number / number of mobile phone users).

\section{Methods}
\begin{figure*}[t!]
	\centering
	\includegraphics[width=0.75\linewidth]{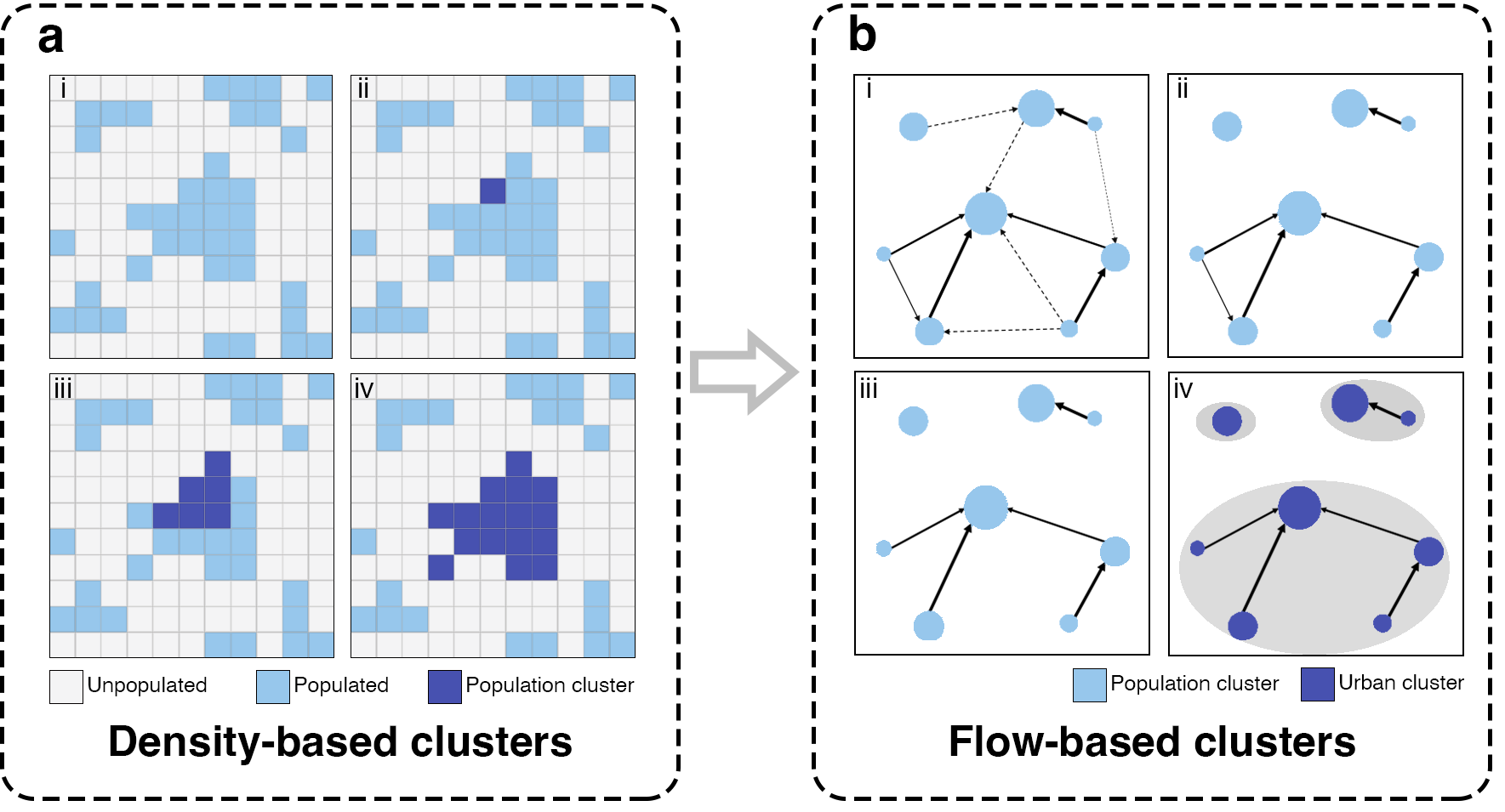}
	\caption{The method. a) Spatially contiguous cells with values larger than the population density threshold are aggregated into population clusters. b)  Population clusters with flow ratios greater than the flow ratio threshold are merged into urban clusters. In panel (b), each node corresponds to a population cluster, and each edge corresponds to the flow ratio from one population cluster to another. Line widths are plotted according to the flow ratio. Solid lines show the edges with weights exceeding the threshold, while dotted lines show the edges with weights less than the threshold.}
	\label{fig:Flowchart}
\end{figure*}

To construct multi-level urban clusters, we take the following steps (Fig. \ref{fig:Flowchart}). First, we merge the spatially adjacent populated grids into clusters at different population density thresholds. We refer to these clusters as population clusters. Second, we merge these population clusters into urban clusters by different flow ratio thresholds. Similar to the concept of FUAs, an urban cluster is a combination of populated areas with strong commuting connections. Finally, we use the percolation theory to analyze the urban clusters under different population density and flow ratio thresholds with the density-flow phase diagram. 

\subsection{Clustering by population density thresholds}
For a given population density threshold $D_*$, we label the cells with values greater than $D_*$ as populated cells and then aggregate geographically contiguous populated cells into population clusters using the City Clustering Algorithm \citep{rozenfeld_laws_2008, rozenfeld_area_2011}. Specifically, we start with any unprocessed populated cell and iteratively add its neighboring populated cells (eight nearest neighbors) until the neighbors of all cells in the cluster are either non-populated cells or within the cluster. Then we aggregate another population cluster until all populated cells belong to a specific cluster. The area and population of each population cluster is the sum of the areas and populations of all cells within that population cluster. In addition, each population cluster should meet certain area and population size criteria. Here, we set the minimum area $A_*$ of the population cluster to $10km^2$ (for reference, the land area of Sansha, the smallest city in China, is about $20km^2$). We also set the minimum area to 5, 15, 20, and 25 $km^2$, and the results are robust (Table A.1); see Appendix for details.

\subsection{Clustering by flow ratio thresholds}
We group the detected population clusters into urban clusters according to different flow ratio thresholds. For given population clusters $i$ and $j$, we measure the commuting flow $P_{ij}$ whose origin is located in cluster $i$ and destination is located in cluster $j$. The flow is then normalized as follows: $F_{ij}=P_{ij}/\sum^{N}_{j=0}{P_{ij}}$. For a given flow ratio threshold $F_*$ and any two population clusters, a link is established between the two clusters if the commuting ratio in either direction exceeds the threshold $F_*$. Besides, if a population cluster has multiple commuting flows to other population clusters that exceed the threshold $F_*$, the cluster is linked only to the cluster with the largest flow ratio. Finally, the linked population clusters are merged together to construct an urban cluster. This process is applied only once for all population clusters in parallel, and the merged clusters will not be iterated again. The area and population of each urban cluster is the sum of the areas and populations of all population clusters within that urban cluster. 

\subsection{Percolation analysis on urban clusters}
We analyze the urban clusters under different density-flow thresholds using the percolation theory. The percolation theory studies the transition of many small clusters merged into a large spanning cluster at the critical point \citep{christensen_complexity_2005}. Around the critical point, cluster systems exhibit power-law critical phenomena that also exist in urban systems (e.g., the size distribution of cities). Therefore, the percolation theory has been widely used to model urban growth \citep{makse_modelling_1995, makse_modeling_1998} and quantify urban clusters \citep{arcaute_cities_2016, cao_quantifying_2020}. Here, we iterate over all possible population density thresholds (from non-zero to very dense) and obtain the population clusters under each density threshold. Then, we traverse the flow ratio thresholds from $0\%$ to $100\%$ and merge the population clusters into urban clusters under each flow threshold, where $0$ means that any two population clusters with interactions are merged together, and $100\%$ means that no population clusters are merged since the flow ratio of two clusters cannot exceed $100\%$. Finally, we obtain many collections of urban clusters, each of which corresponds to a specific population density threshold and a specific flow ratio threshold. Similar to the percolation model, we calculate the area of the largest urban cluster (ranked by land area) as a function of the density and flow thresholds and construct the density-flow phase diagram. 

\section{Results}
\subsection{Density-flow phase diagram}

\begin{figure*}[t!]
	\centering
	\includegraphics[width=0.9\linewidth]{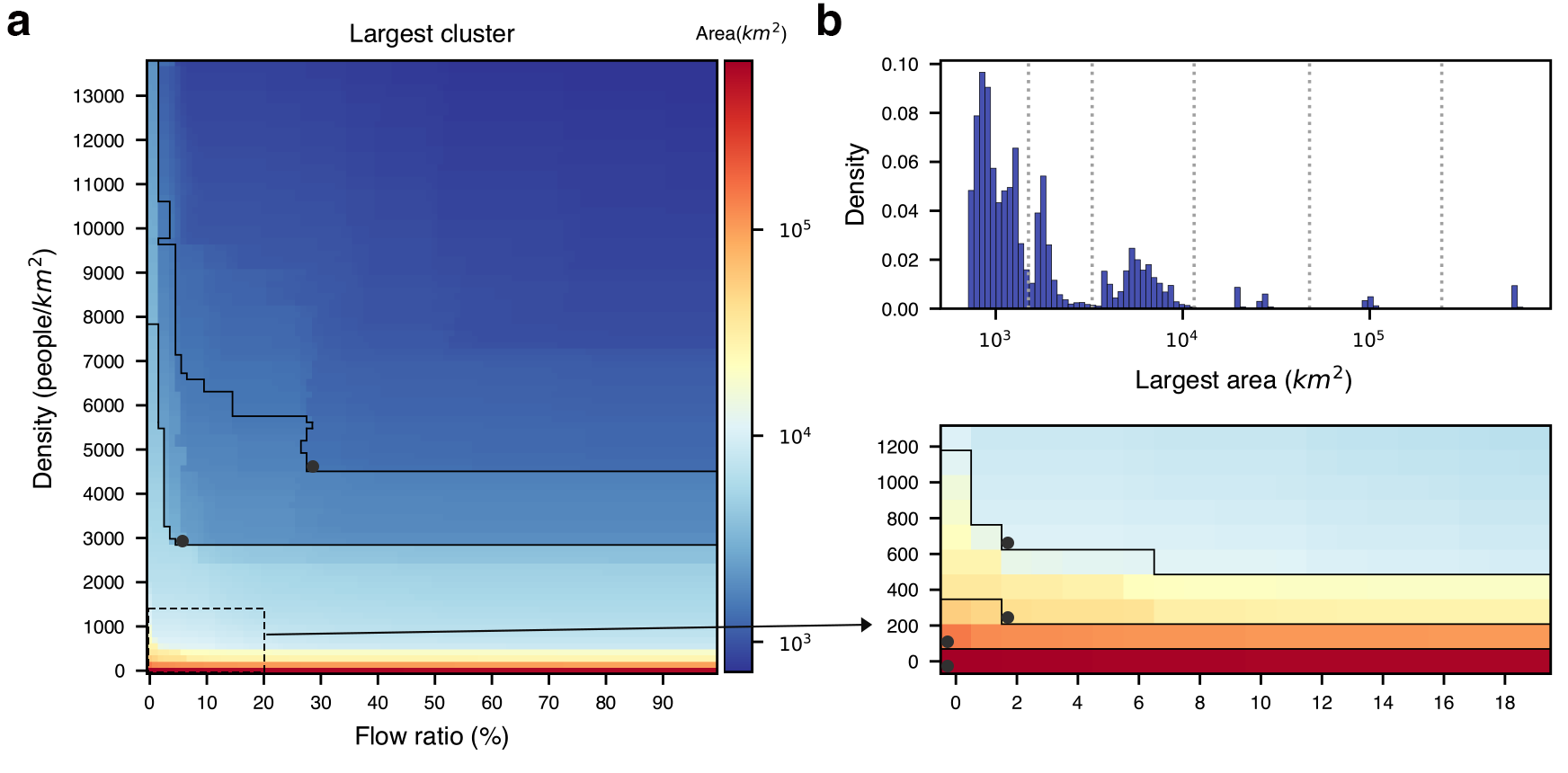}
	\caption{a) The density-flow phase diagram. Each cell in the diagram indicates the size of the largest urban cluster at the corresponding density and flow thresholds. Solid lines indicate the boundaries of each phase. b) Histogram of the largest cluster areas of all urban systems. We observe clear breakpoints that are demarcated by dotted lines.}
	\label{fig:Phase}
\end{figure*}

\begin{table*}[t!]
	\centering
	\caption{Statistics on the urban clusters from Level 1 to Level 6}
		\begin{tabular}{lccccccc}
			\colrule
			Level\quad{} & $D_{*}$ (people/$km^2$)\quad{} & $F_{*}$\quad{} & $N_{urban}$\quad{} & $A_{total}$ ($km^2$)\quad{} & $A_{largest}$ ($km^2$)\quad{} & $P_{total}$ (people)\quad{} & $P_{largest}$ (people)\\  
			\colrule
	 		1 & $\textgreater$ 0 & $\textgreater$ 0\% & 115 & $1.08 \times 10^6$ & $6.60 \times 10^5$ & $1.36 \times 10^9$ & $7.65 \times 10^8$ \\ \cmidrule{1-8}
	 		2 & $\textgreater$ 140 & $\textgreater$ 0\% & 138 & $4.19 \times 10^5$ & $1.45 \times 10^5$ & $1.28 \times 10^9$ & $3.19 \times 10^8$ \\ \cmidrule{1-8}
	 		3 & $\textgreater$ 280 & $\textgreater$ 2\% & 191 & $2.66 \times 10^5$ & $4.26 \times 10^4$ & $1.21 \times 10^9$ & $1.68 \times 10^8$ \\ \cmidrule{1-8}
	 		4 & $\textgreater$ 700 & $\textgreater$ 2\% & 193 & $1.27 \times 10^5$ & $1.07 \times 10^4$ & $1.10 \times 10^9$ & $1.42 \times 10^8$ \\ \cmidrule{1-8}
	 		5 & $\textgreater$ 2,940 & $\textgreater$ 5\% & 349 & $4.89 \times 10^4$ & $3.04 \times 10^3$ & $9.01 \times 10^8$ & $6.47 \times 10^7$ \\ \cmidrule{1-8}
	 		6 & $\textgreater$ 4,620 & $\textgreater$ 28\% & 876 & $3.74 \times 10^4$ & $1.43 \times 10^3$ & $8.27 \times 10^8$ & $4.00 \times 10^7$ \\ 
			\colrule
		\end{tabular}
	\label{tab:Phase}
\end{table*}

We rank urban clusters under each density-flow threshold by land area and obtain the phase diagram (Fig. \ref{fig:Phase}a). The color of each grid cell in Fig. \ref{fig:Phase}a indicates the area of the largest urban cluster under the specified population density-flow thresholds. From bottom left to top right, urban clusters gradually transition from sparsely populated, weakly connected urban areas to densely populated, strongly connected urban areas as the thresholds increase. Remarkably, by transforming Fig. \ref{fig:Phase}a into a histogram of the largest cluster areas under the different density-flow thresholds, we observe $6$ groups (Fig. \ref{fig:Phase}b). Using the breakpoints (vertical dashed lines) in Fig. \ref{fig:Phase}b, we further divide the density-flow phase diagram into $6$ phases (solid lines in Fig. \ref{fig:Phase}a), with the cluster systems remaining stable within each phase. 

As the population density threshold and the flow ratio threshold increase, the largest urban cluster in the system remains stable until the urban system transitions to a lower-level spatial scale. Thus, we regard 6 phases as 6 spatial levels of one urban system. Table \ref{tab:Phase} shows the population density threshold $D_{*}$ (people/$km^2$), the flow ratio threshold $F_{*}$ (\%), the number of urban clusters $N_{urban}$, the total area $A_{total}$ ($km^2$) and the total population $P_{total}$ (people) of the urban clusters, and the area $A_{largest}$ ($km^2$) and the population $P_{largest}$ (people) of the largest cluster in each level from Level 1 to Level 6.

\subsection{Urban clusters at different levels}
\begin{figure*}
	\centering
	\includegraphics[width=0.80\linewidth]{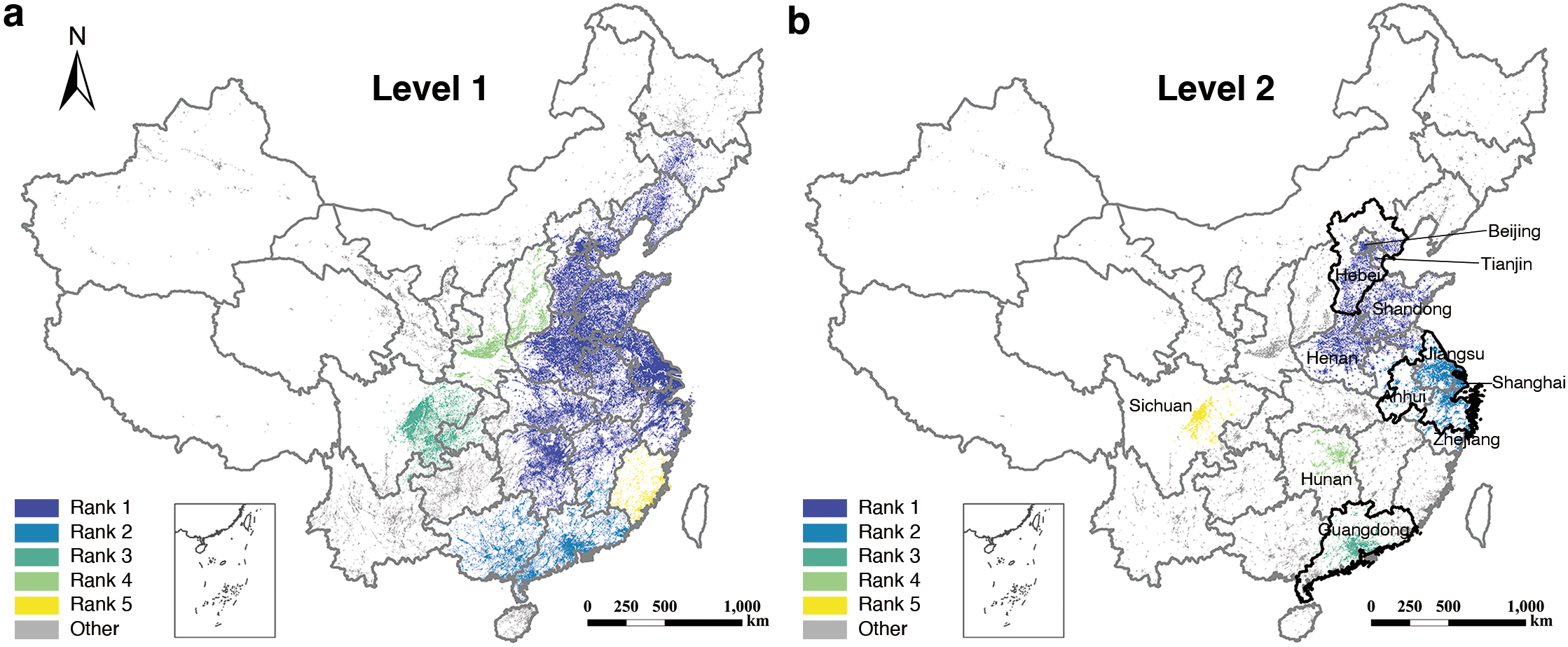}
	\caption{Urban clusters at Level 1 (a) and Level 2 (b). The top $5$ largest clusters are colored by rank. The black lines show the area of Beijing-Tianjin-Hebei, the Yangtze River Delta, and the Pearl River Delta.}
	\label{fig:Phase12}
\end{figure*}

\begin{figure*}
	\centering
	\includegraphics[width=0.8\linewidth]{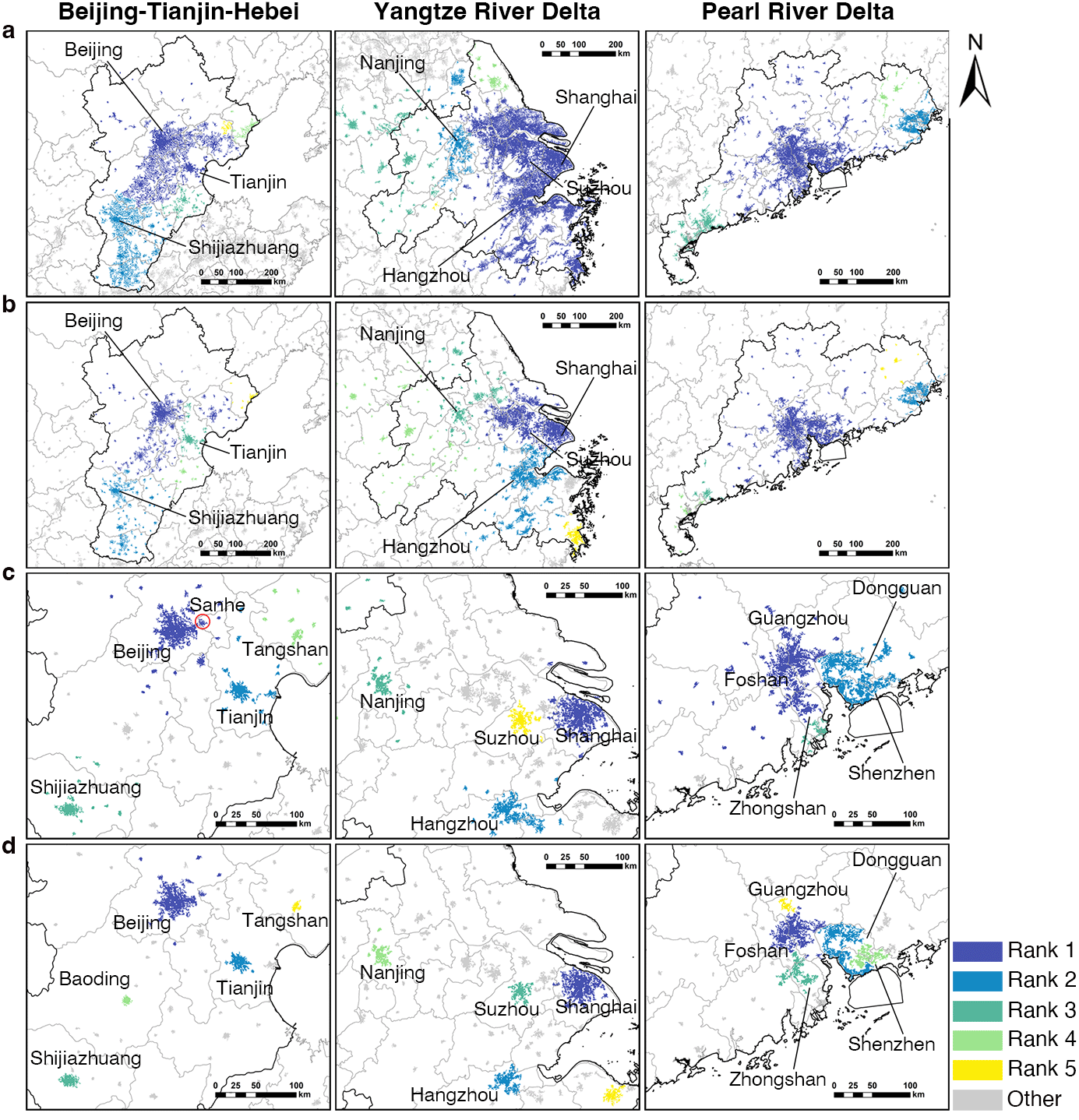}
	\caption{a-d) Urban clusters in Beijing-Tianjin-Hebei, the Yangtze River Delta, and the Pearl River Delta at Levels 3 (a), 4 (b), 5 (c), and 6 (d), respectively. The top $5$ largest clusters in each region are colored by rank.}
	\label{fig:Phase34567}
\end{figure*}

We next investigate the spatial distributions of the urban clusters at different levels. Fig. \ref{fig:Phase12} shows the urban clusters at Level 1 and Level 2, and Fig. \ref{fig:Phase34567} shows the urban clusters from Level 3 to Level 6 in Beijing-Tianjin-Hebei, the Yangtze River Delta, and the Pearl River Delta (mapped with black lines in Fig. \ref{fig:Phase12}b).

Level 1 corresponds to the non-zero population density-flow thresholds; that is, areas with a population $\textgreater 0$ are marked as populated areas, and any two clusters with flows $\textgreater 0$ are linked together. At Level 1, the areas of the top $5$ largest clusters account for $87\%$ of the total area of all clusters. The largest urban cluster has an area of $6.60 \times 10^5 km^2$ and a population of $7.65 \times 10^8$, including most of Central and East China and parts of North and Northeast China. The $2$nd and $5$th largest urban clusters cover South China. The $3$rd and $4$th largest urban clusters correspond to the major urban areas in Southwest and Northwest China, respectively. 

Level 2 corresponds to the thresholds of $140$ people per $km^2$ and a 0\% flow ratio. Level-2 clusters represent the development poles in each region. The largest urban cluster covers most of Hebei Province, Henan Province, and Shandong Province, and has an area of $1.45 \times 10^5 km^2$ and a population of $3.19 \times 10^8$. Beijing, the capital of China, is within the cluster. The $2$nd largest urban cluster covers Jiangsu Province, Anhui Province and Zhejiang Province and includes the Yangtze River Delta; the $3$rd cluster covers Guangdong Province and includes the Pearl River Delta; the $4$th and $5$th clusters correspond to Hunan Province and Sichuan Province, respectively (Fig. \ref{fig:Phase12}b).

Level 3 corresponds to the thresholds of $280$ people per $km^2$ and a 2\% flow ratio, and Level 4 corresponds to the thresholds of $700$ people per $km^2$ and a 2\% flow ratio. The urban clusters at these two levels are consistent with the concept of urban agglomeration, which is the spatial form of several highly integrated cities \citep{fang_urban_2017}. The largest urban cluster at Level 3 is located in the Yangtze River Delta, including Shanghai and major cities in southern Jiangsu Province and northern Zhejiang Province. At Level 4, Hangzhou and its surrounding cities are separated from this cluster. The largest urban cluster at Level 4 is located in the Pearl River Delta, corresponding to the Greater Bay Area, and mainly includes Guangzhou and Shenzhen (both cities with a population of more than 10 million). The Yangtze River Delta and the Pearl River Delta are all well-developed regions in China.

Level 5 corresponds to the thresholds of $2,940$ people per $km^2$ and a 5\% flow ratio. At Level 5, each urban cluster has a dense core urban area, as well as surrounding areas that are closely connected to the core, which is similar to the definition of an FUA. However, instead of using pre-defined parameters (population density and flow ratio thresholds), our method obtains the thresholds objectively through percolation. The number of Level-5 urban clusters (349) is similar to the number of prefecture-level cities (333) in China. As shown in Fig. \ref{fig:Phase34567}, one prefecture-level city corresponds to one urban cluster in most regions. Yet, in some well-developed regions, intensive socio-economic connections make urban clusters cross administrative boundaries. For example, Sanhe (the red circle in Fig. \ref{fig:Phase34567}c) and Beijing are merged together due to the large number of commuting, although Beijing and Sanhe are within different administrative city boundaries.  

Level 6 corresponds to the population density threshold of $4,620$ people per $km^2$ and the flow ratio threshold of 28\%. Due to the high flow ratio threshold, most urban clusters in this level consist of one single population cluster. Compared to the administrative boundaries, the obtained urban clusters mainly correspond to core urban areas within prefecture-level cities. There are also some urban clusters that cross the boundaries of prefecture-level cities. For example, the central bodies of western Shenzhen and Dongguan are geographically continuous with high population density, forming an urban cluster across the boundaries, while eastern Shenzhen is not adjacent to this urban cluster due to geographical barriers (e.g., mountains and forests), and the flow ratio is also not high enough, forming another urban cluster. As a validation, we compare the largest urban cluster (Beijing) detected by our method to the official released built-up area data in 2015. Our results show that Beijing at Level 6 has an area of $1,381 km^2$, while the official number is $1,401km^2$, and the two values are very close. 

We also perform a sensitivity test by applying our method to Guangdong Province in China. Although fewer levels are derived, the density-flow phase diagram and the derived urban clusters at different levels are similar; see Fig. A.1-A.2 in the Appendix.

\subsection{Zipf's law and allometric growth}
To investigate the relationship between area and population in the urban clusters, we test Zipf's law and allometric growth. Zipf's law and allometric growth reflect the self-organized properties of urban complex systems and have been widely tested across different countries in different periods \citep{soo_zipfs_2005, bettencourt_growth_2007, bettencourt_origins_2013}. Zipf's law states that when cities are ranked by population size, the population size $S$ and the rank $R$ of cities satisfy $S=1/R$. This can also be expressed as the probability density function (PDF) of urban cluster sizes following a power-law distribution where $P(S)\sim S^{-\alpha}$ and $\alpha = 2$ \citep{zipf_human_1949, gabaix_zipfs_1999}. Allometric growth refers to the sub-linear power-law relationship between the population $P$ and the urban area $A$, which can be expressed as $A=P^\beta$ with $\beta \approx 0.6-0.8$ \citep{nordbeck_urban_1971, bettencourt_origins_2013}. Allometric growth represents economies of scale, meaning that larger urban clusters can result in more efficient land use per capita, since the population is more concentrated \citep{bettencourt_growth_2007}.

\begin{figure*}[t]
	\centering
	\includegraphics[width=0.75\linewidth]{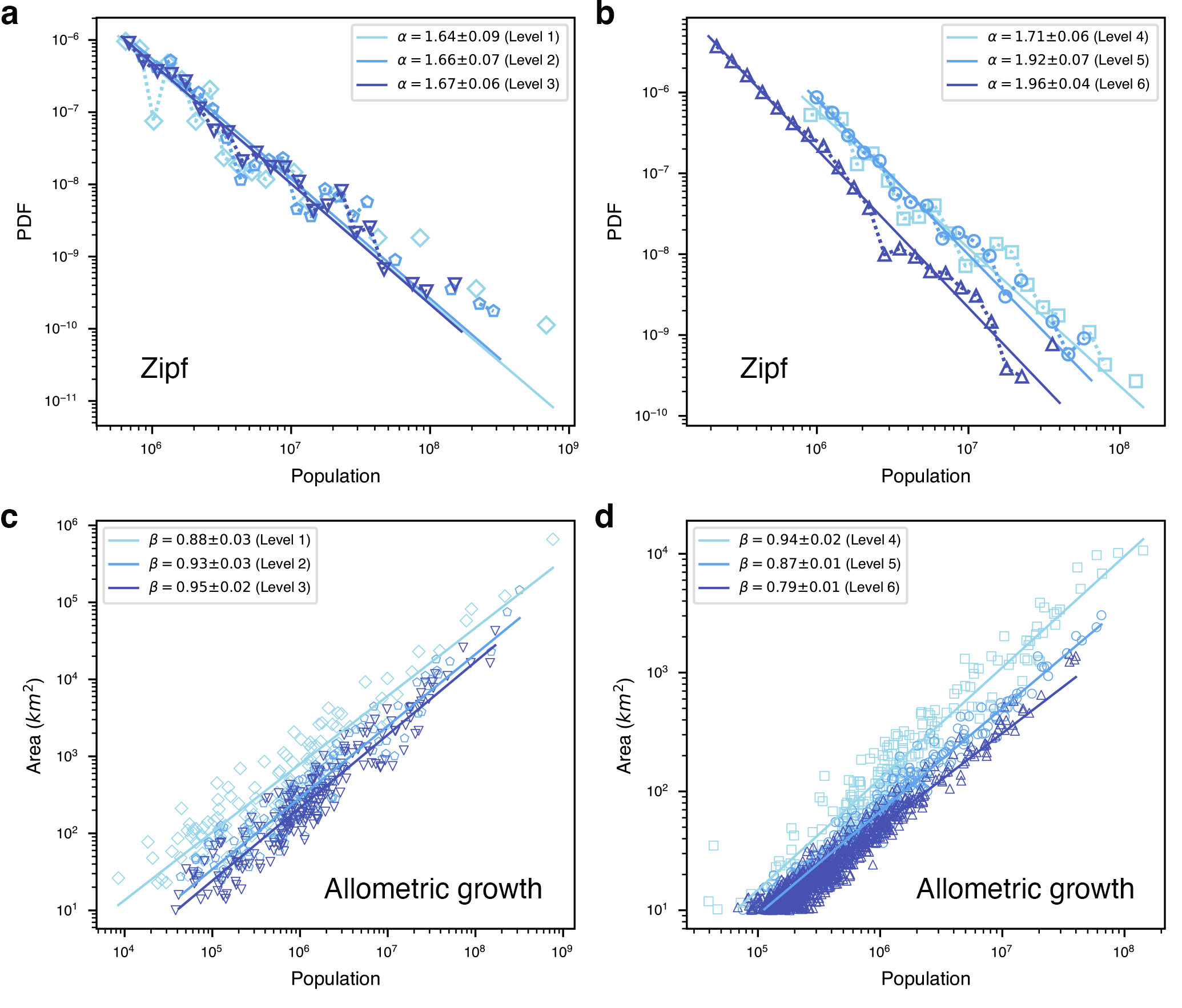}
	\caption{a-b) The PDFs of the population sizes of the urban clusters from Level 1 to Level 6. Solid lines represent the power-law fitting results. c-d) The sub-linear relationship between the populations and the areas of the urban clusters from Level 1 to Level 6. Solid lines represent the fitting results. $\pm$ represents one standard deviation.}
	\label{fig:Zipf}
\end{figure*}

Figure \ref{fig:Zipf}a-b shows that the population sizes of the urban clusters from Level 1 to Level 6 all follow power-law distributions. We further use the method proposed by \citet{alstott_powerlaw_2014} to fit the data, and the power-law exponents $\alpha$ for Level 5 and Level 6 are around $2$, which means that Zipf's law holds well at these two levels. However, the exponents $\alpha$ for Levels 1-4 deviate from $2$, indicating that Zipf's law does not hold at higher levels. It should be noted that there is a long-standing debate on whether Zipf's law holds for urban clusters at different scales \citep{berry_city_2012}, and Zipf's law is not a validation of the derived urban clusters. As a complement, we demonstrate that Zipf's law does not hold on all spatial scales; it only holds after Level 5. Interestingly, as shown in the previous section, the Level-5 urban clusters have similar spatial scales to the prefecture-level cities, indicating that our results are consistent with previous studies on the empirical test of Zipf's law \citep{gabaix_zipfs_1999, soo_zipfs_2005}.

Figure \ref{fig:Zipf}c-d shows the power-law relationship between the populations and the areas of the urban clusters at different levels. The exponents $\beta$ are all $<1$, indicating the existence of allometric growth. However, similar to Zipf's law, only after Level 5, the exponents begin to approach the previous theoretical and empirical values \citep{bettencourt_origins_2013}. This finding also adds evidence that the allometric growth depends on the definition of urban clusters \citep{berry_city_2012, oliveira_large_2014}. 

\section{Discussion}
We propose a bottom-up approach to construct multi-level urban clusters using population distribution and flow data. The density-flow thresholds are derived according to the critical characteristics of urban systems based on the percolation theory. We reveal six levels of urban clusters from the density-flow phase diagram, and show that Zipf's law holds only for urban clusters after Level 5, providing empirical evidence for the spatially dependent nature of urban laws. Regarding applications, our method does not rely on any pre-defined geographical units and requires only the density-flow thresholds obtained by percolation, allowing us to efficiently derive multi-level urban clusters. 

The prevalence of mobile phones allows us to obtain population distribution and flow data at a lower cost and faster update rate. Thus, our approach can be easily applied to countries that have mobile phone data but do not have recent census/survey data (e.g., some developing countries). Even for those regions where mobile phone data are unavailable, we could also consider using human mobility models to infer population flows from population distributions \citep{barbosa_human_2018}, which are usually easy to obtain, such as the WorldPop dataset \citep{tatem_worldpop_2017}.

Some extensions can be explored in further studies. First, in this paper, we use commuting flows to measure the socio-economic connections between areas. However, commuting accounts for only approximately 1/3 of all human movements \citep{ding_investigating_2017}, and a large number of non-commuting movements also reflect the socio-economic connections. The impact of different types of flow interactions on quantifying urban clusters is worth exploring. Second, the movements of individuals vary substantially across different periods. For example, resident commuting dropped significantly during the COVID-19 pandemic \citep{kraemer_effect_2020}. This might lead to changes in the boundaries of flow-based urban clusters, and these temporally dynamic boundaries are worth exploring in our future work.

\section{Data and codes availability}
The dataset and codes to reproduce the results of this paper are available via \url{https://github.com/caowenpu56/densityflow}.

\end{document}